\documentclass[aps, showpacs, showkeys,nofootinbib,floatfix]{revtex4}

\usepackage{amssymb}
\usepackage{amsmath}
\usepackage{graphicx}

\begin{document}

\title{Constraints on Kinematic Model from Recent Cosmic Observations: SN Ia, BAO and Observational Hubble Data}

\author{Lixin Xu\footnote{Corresponding author}}
\email{lxxu@dlut.edu.cn}
\author{Wenbo Li}
\author{Jianbo Lu}

\affiliation{Institute of Theoretical Physics, School of Physics \&
Optoelectronic Technology, Dalian University of Technology, Dalian,
116024, P. R. China}

\begin{abstract}
In this paper, linear first order expansion of deceleration
parameter $q(z)=q_0+q_1(1-a)$ ($M_1$), constant jerk $j=j_0$ ($M_2$)
and third order expansion of luminosity distance ($M_3$) are
confronted with cosmic observations: SCP $307$ SN Ia, BAO and
observational Hubble data (OHD). Likelihood is implemented to find
the best fit model parameters. All these models give the same
prediction of the evolution of the universe which is undergoing
accelerated expansion currently and experiences a transition from
decelerated expansion to accelerated expansion. But, the transition
redshift depends on the concrete  parameterized form of the model
assumed. $M_1$ and $M_2$ give value of transition redshift about
$z_t\sim 0.6$. $M_3$ gives a larger one, say $z_t\sim 1$. The
$\chi^2/dof$ implies almost the same goodness of the models. But,
for its badness of evolution of deceleration parameter at high
redshift $z>1$, $M_3$ can not be reliable. $M_1$ and $M_2$ are
compatible with $\Lambda$CDM model at the $2\sigma$ and $1\sigma$
confidence levels respectively. $M_3$ is not compatible with
$\Lambda$CDM model at $2\sigma$ confidence level. From $M_1$ and
$M_2$ models, one can conclude that the cosmic data favor a
cosmological model having $j_0<-1$.
\end{abstract}

\pacs{Added}

 \hfill TP-DUT/2009-05

\maketitle

\section{Introduction}

The expansion of the universe or kinematics of the universe is
described by the expansion rate $H=\dot{a}/a$, a dimensionless
parameter $q=-a\ddot{a}/\dot{a}^2$, named deceleration parameter and
jerk parameter $j=-\dddot{a}a^3/(a\dot{a}^3)$, where $a$ is the
scale factor in Friedmann-Roberson-Walker (FRW) metric. The front
negative sign of $q$ is added to obtain a positive parameter when
one considers a decelerated expansion universe dominated by matter
fields with attractive force, for example dark matter dominated
universe with $q=1/2$. However, the observations of Type Ia
Supernovae (SN Ia) from two teams \cite{ref:Riess98,ref:Perlmuter99}
imply that our universe is undergoing an accelerated expansion at
present. Whereafter, this result is confirmed by the observations
from WMAP \cite{ref:Spergel03,ref:Spergel06} and Large Scale
Structure survey \cite{ref:Tegmark1,ref:Tegmark2}. Then at present,
the deceleration parameter would be a negative number. In fact, to
explain the accelerated expansion of the universe, a flood of
various cosmological models have been explored, please see
\cite{ref:reviews} for recent reviews.

In the literatures, two approaches have been taken into accounts.
One is that an extra energy component, dubbed dark energy, is
introduced, e.g. cosmological constant, quintessence
\cite{ref:quintessence1,ref:quintessence2,ref:quintessence3,ref:quintessence4},
phantom \cite{ref:phantom} and quintom \cite{ref:quintom}, etc. The
other is that the accelerated expansion of the universe is due to
the modification of the gravity theory at large scale, e.g. modified
gravity theory, Brans-Dicke theory and higher dimensional theory,
etc. These correspond to the dynamics of the universe. In general,
different dark energy models would predict different expansion
histories of the universe, for example quintessence dark energy
model with various scaler potentials \cite{ref:Sahni} will have
different values of the transition redshift $z_T$ from decelerated
expansion to accelerated expansion. These different models would
have different dynamics. However, a kinematic approach will be held
regardless of the underlying cosmic dynamics
\cite{ref:kinematic1,ref:kinematic2}. In particular, the jerk
parameter can provide us the simplest approach to search for
departures from the cosmic concordance model, for its constant value
$j=-1$ for cosmological constant. This approach is called
cosmography \cite{ref:cosmography,ref:ST2006}, cosmokinetics
\cite{ref:cosmokinetics}, or Friedmannless cosmology
\cite{ref:Friedmannless1,ref:Friedmannless2}.

Some explore the accelerated expansion of the universe by using
different parameterized forms of deceleration parameters $q(z)$
\cite{ref:gongpq,ref:xupq,ref:BCK2004} in the so-called model
independent way, for examples constant model $q=constant$, linear
model with variable $z$ ($q(z)=q_0+q_1 z$) and linear model with
variable $a$ ($q(a)=q_0+q_1(1-a)$), etc. Recently, the authors of
\cite{ref:kinematic2,ref:Friedmannless2} investigated constraints on
some kinematic models by employing a Bayesian marginal likelihood
analysis based on the cosmic observations from supernovae type Ia
with extent including the jerk parameter $j$: the third order
contribution in the expansion for kinematic luminosity distance in
terms of the redshift $z$.

As known, the degeneracies between the parameters will be broken
when more different observational data added in constraint to
models. In this work, we will go further by including BAO and the
observational Hubble data in examining the case for the jerk
parameter (constant and variable).

\section{Kinematic Models}

The metric of a flat FRW cosmological model containing dark matter
and dark energy is written as
\begin{equation}
ds^2=-dt^2+a^2(t)dx^2.
\end{equation}
where $a(t)$ is the scale factor, which describes the whole history
of the universe evolution, and has the relations with redshift $z$
in terms of $a=(1+z)^{-1}$ ($a_0=1$ is normalized). The Hubble
parameter
\begin{equation}
H\equiv \frac{\dot{a}}a,\label{eq:hubble}
\end{equation}
and deceleration parameter
\begin{equation}
q\equiv -\frac{1}{H^2}\frac{\ddot{a}}{a}= \frac 12
(1+z)\frac{[H(z)^2]'}{H(z)^2}-1.\label{eq:dec}
\end{equation}
are defined as the rate of expansion and accelerated expansion. By
using the relation $a_0/a=1+z$ and the relations between $H$ and
$q$, {\it i.e.}, one can rewrite Eq. (\ref{eq:dec}) in its
integration form
\begin{equation}
H(z)=H_0\exp\left[\int_{0}^{z}\left[1+q(u)\right]d\ln(1+u)\right].\label{eq:q_int}
\end{equation}
Similarly, the jerk parameter is defined as
\begin{equation}
j\equiv -\frac{1}{H^3} \frac{\dot{\ddot{a}}}{a}=-\left[\frac 12
(1+z)^2
\frac{[H(z)^2]''}{H(z)^2}-(1+z)\frac{[H(z)^2]'}{H(z)^2}+1\right].\label{eq:jerk}
\end{equation}
Easily, one can find that the deceleration parameter and jerk
parameter have the relations
\begin{equation}
j=-\left[ q + 2q^{2}+(1 + z)\frac{dq}{dz}\right],
\end{equation}
which will be used when the parameterized forms of $q(z)$ are giong
to be tested. By using these definitions Eq. (\ref{eq:hubble}), Eq.
(\ref{eq:dec}) and Eq. (\ref{eq:jerk}), one can describe the recent
cosmic expansion with their current values
\begin{equation}
a(t)= 1+H_0(t-t_0) - \frac{1}{2}q_0 H_0^2(t-t_0)^2 - \frac{1}{3!}j_0
H_0^3(t-t_0)^3 + {\cal O}[(t-t_0)^4],\label{eq:a_exp}
\end{equation}
from which the luminosity distance can be expanded as
\cite{ref:Weinb72}
\begin{equation}
d_L(z)=\frac{c}{H_0}\left[z +
\frac{1}{2}(1-q_0)z^2-\frac{1}{6}(1-q_0-3q_0^2-j_0)z^3\right] +{\cal
O}(z^4).\label{eq:dL_exp}
\end{equation}
Also, one can find the relations of Hubble parameter $H(z)$ and
luminosity distance $d_{L}(z)$ \cite{ref:Weinb72}
\begin{eqnarray}
H^{-1}(z)&=&-(1+z)\frac{dt}{dz}\nonumber\\
&=&\frac{d}{dz}\left[(1+z)^{-1}d_{L}(z)\right]
\end{eqnarray}
which will be useful when the observational Hubble data are used as
cosmic observation constraint.

The basic aim in this paper is to examine some simple kinematic
models for the cosmic expansion based on specific parameterizations
for $q(z)$ in Eq. (\ref{eq:dec}) (variable jerk parameter), a
constant jerk parameter and their comparison with the expansion
(\ref{eq:a_exp}).

The first model, $M_1$, is given by linear expansion of the scale
factor $a$, $q(a)=q_0+q_1(1-a)$ in terms of $a$, which can be
rewritten in the terms redshift $z$, $q(z)=q_0+q_1z/(1+z)$. The
second model, $M_2$, is a constant jerk parametrization model,
$j(z)=j_0$, for detecting the departure from the flat $\Lambda$CDM
scenario, for which $j(z)=j_0=-1$. Model $M_3$ is the expansion
(\ref{eq:a_exp}), which has as free parameters $q_0$ and $j_0$. In
the Appendix \ref{sec:append}, see also \cite{ref:kinematic2}, one
can find the basic analytical expressions between the Hubble
parameter $H(z)$, deceleration parameter $q(z)$ and jerk parameter
$j(z)$. One can take this paper as a generalization and complement
to \cite{ref:kinematic2} where only the $307$ SN Ia data points are
used. Here, the BAO and OHD datastes are also included as useful
cosmic constraints. One would notice that all of them do not include
$\Omega_{m}$ explicitly.

\section{Cosmic observation data sets and statistical results}


\subsection{SN Ia}
We constrain the parameters with the Supernovae Cosmology Project
(SCP) Union sample including $307$ SN Ia \cite{ref:SCP}, which
distributed over the redshift interval $0.015\le z\le 1.551$.
Constraints from SN Ia can be obtained by fitting the distance
modulus $\mu(z)$
\begin{equation}
\mu_{th}(z)=5\log_{10}(D_{L}(z))+\mu_{0},
\end{equation}
where, $D_{L}(z)$ is the Hubble free luminosity distance $H_0
d_L(z)/c$ and
\begin{eqnarray}
d_L(z)&=&c(1+z)\int_{0}^{z}\frac{dz^{\prime}}{H(z^{\prime})}\\
\mu_0&\equiv&42.38-5\log_{10}h,
\end{eqnarray}
where $H_0$ is the Hubble constant which is denoted in a
re-normalized quantity $h$ defined as $H_0 =100 h~{\rm km ~s}^{-1}
{\rm Mpc}^{-1}$. The observed distance moduli $\mu_{obs}(z_i)$ of SN
Ia at $z_i$ is
\begin{equation}
\mu_{obs}(z_i) = m_{obs}(z_i)-M,
\end{equation}
where $M$ is their absolute magnitudes.

For SN Ia dataset, the best fit values of parameters in a model can
be determined by the likelihood analysis is based on the calculation
of
\begin{eqnarray}
\chi^2(p_s,m_0)&\equiv& \sum_{SNIa}\frac{\left[
\mu_{obs}(z_i)-\mu_{th}(p_s,z_i)\right]^2} {\sigma_i^2} \nonumber\\
&=&\sum_{SNIa}\frac{\left[ 5 \log_{10}(D_L(p_s,z_i)) - m_{obs}(z_i)
+ m_0 \right]^2} {\sigma_i^2}, \label{chi2}
\end{eqnarray}
where $m_0\equiv\mu_0+M$ is a nuisance parameter (containing the
absolute magnitude and $H_0$) that we analytically marginalize over
\cite{ref:SNchi2},
\begin{equation}
\tilde{\chi}^2(p_s) = -2 \ln \int_{-\infty}^{+\infty}\exp \left[
-\frac{1}{2} \chi^2(p_s,m_0) \right] dm_0 \; ,
\label{chi2_marginalization}
\end{equation}
to obtain
\begin{equation}
\tilde{\chi}^2 =  A - \frac{B^2}{C} + \ln \left(
\frac{C}{2\pi}\right) , \label{chi2_marginalized}
\end{equation}
where
\begin{equation}
A=\sum_{SNIa} \frac {\left[5\log_{10}
(D_L(p_s,z_i))-m_{obs}(z_i)\right]^2}{\sigma_i^2},
\end{equation}
\begin{equation}
B=\sum_{SNIa} \frac {5
\log_{10}(D_L(p_s,z_i)-m_{obs}(z_i)}{\sigma_i^2},
\end{equation}
\begin{equation}
C=\sum_{SNIa} \frac {1}{\sigma_i^2} \; .
\end{equation}
The Eq. (\ref{chi2}) has a minimum at the nuisance parameter value
$m_0=B/C$. Sometimes, the expression
\begin{equation}
\chi^2_{SNIa}(p_s,B/C)=A-(B^2/C)\label{eq:chi2SNIa}
\end{equation}
is used instead of Eq. (\ref{chi2_marginalized}) to perform the
likelihood analysis. They are equivalent, when the prior for $m_0$
is flat, as is implied in (\ref{chi2_marginalization}), and the
errors $\sigma_i$ are model independent, what also is the case here.
Obviously, from the value $m_0=B/C$, one can obtain the best-fit
value of $h$ when $M$ is known.

To determine the best fit parameters for each model, we minimize
$\chi^2(p_s,B/C)$ which is equivalent to maximizing the likelihood
\begin{equation}
{\cal{L}}(p_s) \propto e^{-\chi^2(p_s,B/C)/2} .
\end{equation}

\subsection{BAO}
The BAO are detected in the clustering of the combined 2dFGRS and
SDSS main galaxy samples, and measure the distance-redshift relation
at $z = 0.2$. BAO in the clustering of the SDSS luminous red
galaxies measure the distance-redshift relation at $z = 0.35$. The
observed scale of the BAO calculated from these samples and from the
combined sample are jointly analyzed using estimates of the
correlated errors, to constrain the form of the distance measure
$D_V(z)$ \cite{ref:Okumura2007,ref:Eisenstein05,ref:Percival}
\begin{equation}
D_V(z)=\left[(1+z)^2 D^2_A(z) \frac{cz}{H(z)}\right]^{1/3},
\label{eq:DV}
\end{equation}
where $D_A(z)$ is the proper (not comoving) angular diameter
distance which has the following relation with $d_{L}(z)$
\begin{equation}
D_A(z)=\frac{d_{L}(z)}{(1+z)^2}.
\end{equation}
Matching the BAO to have the same measured scale at all redshifts
then gives \cite{ref:Percival}
\begin{equation}
D_{V}(0.35)/D_{V}(0.2)=1.812\pm0.060.
\end{equation}
Then, the $\chi^2_{BAO}(p_s)$ is given as
\begin{equation}
\chi^2_{BAO}(p_s)=\frac{\left[D_{V}(0.35)/D_{V}(0.2)-1.812\right]^2}{0.060^2}\label{eq:chi2BAO}.
\end{equation}

\subsection{OHD}

The observational Hubble data are based on differential ages of the
galaxies \cite{ref:JL2002}. In \cite{ref:JVS2003}, Jimenez {\it et
al.} obtained an independent estimate for the Hubble parameter using
the method developed in \cite{ref:JL2002}, and used it to constrain
the EOS of dark energy. The Hubble parameter depending on the
differential ages as a function of redshift $z$ can be written in
the form of
\begin{equation}
H(z)=-\frac{1}{1+z}\frac{dz}{dt}.
\end{equation}
So, once $dz/dt$ is known, $H(z)$ is obtained directly
\cite{ref:SVJ2005}. By using the differential ages of
passively-evolving galaxies from the Gemini Deep Deep Survey (GDDS)
\cite{ref:GDDS} and archival data
\cite{ref:archive1,ref:archive2,ref:archive3,ref:archive4,ref:archive5,ref:archive6},
Simon {\it et al.} obtained $H(z)$ in the range of $0\lesssim z
\lesssim 1.8$ \cite{ref:SVJ2005}. The observational Hubble data from
\cite{ref:SVJ2005} are list in Table \ref{Hubbledata}.
\begin{table}[htbp]
\begin{center}
\begin{tabular}{c|lllllllll}
\hline\hline
 $z$ &\ 0.09 & 0.17 & 0.27 & 0.40 & 0.88 & 1.30 & 1.43
 & 1.53 & 1.75\\ \hline
 $H(z)\ ({\rm km~s^{-1}\,Mpc^{-1})}$ &\ 69 & 83 & 70
 & 87 & 117 & 168 & 177 & 140 & 202\\ \hline
 $1 \sigma$ uncertainty &\ $\pm 12$ & $\pm 8.3$ & $\pm 14$
 & $\pm 17.4$ & $\pm 23.4$ & $\pm 13.4$ & $\pm 14.2$
 & $\pm 14$ &  $\pm 40.4$\\
\hline
\end{tabular}
\end{center}
\caption{\label{Hubbledata} The observational $H(z)$
data~\cite{ref:SVJ2005,ref:JVS2003}.}
\end{table}

The best fit values of the model parameters from observational
Hubble data \cite{ref:SVJ2005} are determined by minimizing
\begin{equation}
\chi_{Hub}^2(p_s)=\sum_{i=1}^9 \frac{[H_{th}(p_s;z_i)-H_{
obs}(z_i)]^2}{\sigma^2(z_i)},\label{eq:chi2H}
\end{equation}
where $p_s$ denotes the parameters contained in the model, $H_{th}$
is the predicted value for the Hubble parameter, $H_{obs}$ is the
observed value, $\sigma(z_i)$ is the standard deviation measurement
uncertainty, and the summation is over the $9$ observational Hubble
data points at redshifts $z_i$.

\subsection{Statistical Results}

For Gaussian distributed measurements, the likelihood function
$L\propto e^{-\chi^2/2}$, where $\chi^2$ is
\begin{equation}
\chi^2=\chi^2_{SNIa}+\chi^2_{BAO}+\chi^2_{Hub},
\end{equation}
where $\chi^2_{SNIa}$, $\chi^{2}_{BAO}$ and $\chi^2_{Hub}$ are the
ones described in Eq. (\ref{eq:chi2SNIa}), Eq. (\ref{eq:chi2BAO})
and Eq. (\ref{eq:chi2H}) respectively. It is clear that only the
kinematic variables, say $h_0$, $q_0$ and $j_0$, are contained in
all $\chi^2$s equations where $\Omega_{m}$ does not appear
explicitly. Then, the results may not depend on the dynamic
variables, say $\Omega_{m}$, and gravitation theory. Here more
datasets are included to constrain the model parameters than that in
Ref. \cite{ref:kinematic2} where the SCP $307$ is used alone. After
the calculation as described above, the result is listed in Table
\ref{Tab:results}.
\begin{table}[htbp]
\begin{center}
\begin{tabular}{c|cccccc}
\hline Models & $\chi^2_{min}$ & $q_0(1\sigma)$ & $j_0(1\sigma)$ &
$h_0(1\sigma)$ & $z_t(1\sigma)$ & $\chi^2/dof$ \\ \hline \hline
$M_1$ & $326.599$ & $-0.715^{+0.045}_{-0.045}$ &
$-2.196^{+0.254}_{-0.241}$ & $0.716^{+0.053}_{-0.053}$ &
$0.609^{+0.110}_{-0.070}$ & $1.043$
\\ \hline
$M_2$ & $326.442$ & $-0.658^{+0.061}_{-0.057}$ &
$-1.382^{+0.219}_{-0.225}$ & $0.709^{+0.053}_{-0.053}$ &
$0.592^{+0.099}_{-0.064}$ & $1.040$ \\ \hline
$M_3$ & $330.733$ & $-0.461^{+0.031}_{-0.033}$ & $-0.147^{+0.110}_{-0.094}$ & $0.716^{+0.053}_{-0.053}$ & $0.946^{+0.110}_{-0.081}$ & $1.053$ \\
\hline
\end{tabular}
\end{center}
\caption{\label{Tab:results} The calculation results of the models.
The $dof$ denotes the degrees of freedom of the models.}
\end{table}

\begin{figure}[tbh]
\centering
\includegraphics[width=7in]{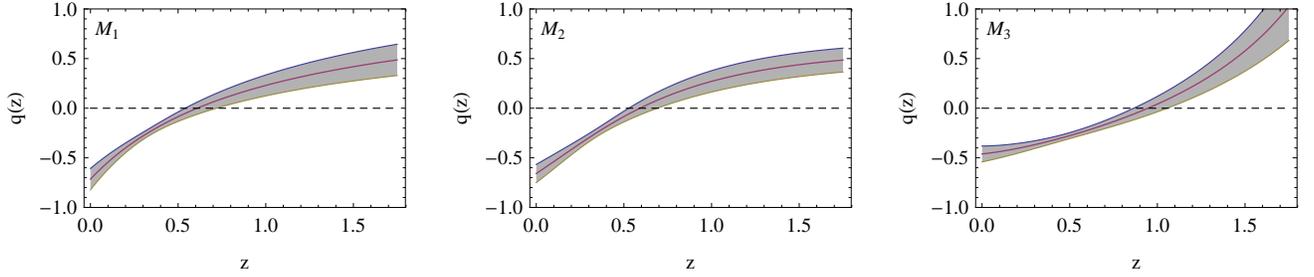}
\caption{The evolutions of deceleration parameters $q(z)$ with
respect to redshift $z$ in $1\sigma$ error regions. The left, center
and right panels correspond to $M_1$ ($q(z)=q_0+q_1z/(1+z)$), $M_2$
($j=j_0$) and $M_3$(expansion of luminosity distance)
respectively.}\label{fig:qz}
\end{figure}

It can be easily seen that $M_1$ and $M_2$ almost give the same
result and have the same goodness in the viewpoint of $\chi^2/dof$.
There the jerk parameter $j_0$ is different: $M_1$ has larger
absolute value than that of $M_2$. However, $M_3$ has smaller
absolute value of $q_0$ and $j_0$ than that of $M_1$ and $M_2$ and
predict larger transition redshift from decelerated expansion to
accelerated expansion. The corresponding evolution curves of
deceleration parameter $q(z)$ with respect to redshift $z$ are
plotted in Fig. \ref{fig:qz}. From the Fig. \ref{fig:qz}, one can
see that $M_1$ and $M_2$ almost give the same evolution history of
the universe. The difference between them is that the $q(z)$ of
$M_2$ is depressed at high redshift than that of $M_1$. However,
$M_3$ almost gives the wrong evolution history of the universe at
high redshift for $q$ approximates to $0.5$ when the epoch of matter
dominated. The reason is simple that the model $M_3$, expansion of
scale factor at low redshift, is not reliable at high redshift. But,
we must notice that $M_3$ is the worst one among these three models
in the viewpoint of $\chi^2/dof$. This is very different from the
results of $M_3$ obtained in Ref. \cite{ref:kinematic2} where SCP
$307$ is used alone. For $\Lambda$CDM model, by using the datastes
of SN Ia, BAO and OHD, we found the best fit values:
$\Omega_{m0}=0.283^{+0.033}_{-0.031}$ and
$h_0=0.732^{+0.044}_{-0.044}$. And the corresponding minimum value
of $\chi^2$ is $\chi^2_{min}=327.246$ ($\chi^2/dof=1.039$). But, it
is clear all the models give the same prediction that the universe
is undergoing accelerated expansion and experience an transition
from decelerated expansion to accelerated expansion. The transition
redshift $z_t$ depends on the concrete parameterized form one
assumed. Here $M_1$ and $M_2$ give the value about $0.6$. However,
the luminosity expansion model $M_3$ give large transition redshift.
From the right panel of Fig. \ref{fig:qz}, one can not take the
expansion model seriously for its bad behavior at relative high
redshift, say $z>1$.

\begin{figure}[tbh]
\centering
\includegraphics[width=6in]{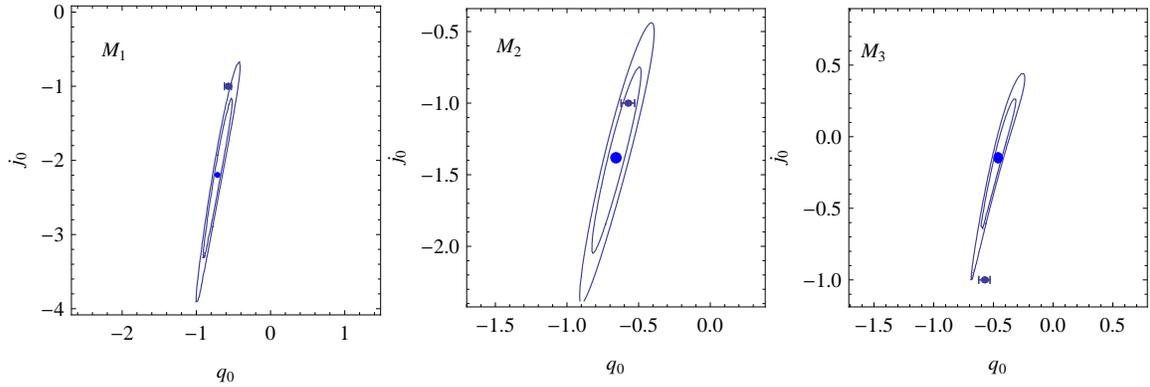}
\caption{The contour plots of $q_0-j_0$ with $1\sigma$ and $2\sigma$
regions. The left, center and right panels correspond to $M_1$
($q(z)=q_0+q_1z/(1+z)$), $M_2$ ($j=j_0$) and $M_3$(expansion of
luminosity distance) respectively, where the dots denote the best
fit values of the parameters. The $\Lambda$CDM model result is
included as a short line segment denoting $1\sigma$
interval.}\label{fig:con_q_j}
\end{figure}

\begin{figure}[tbh]
\centering
\includegraphics[width=6in]{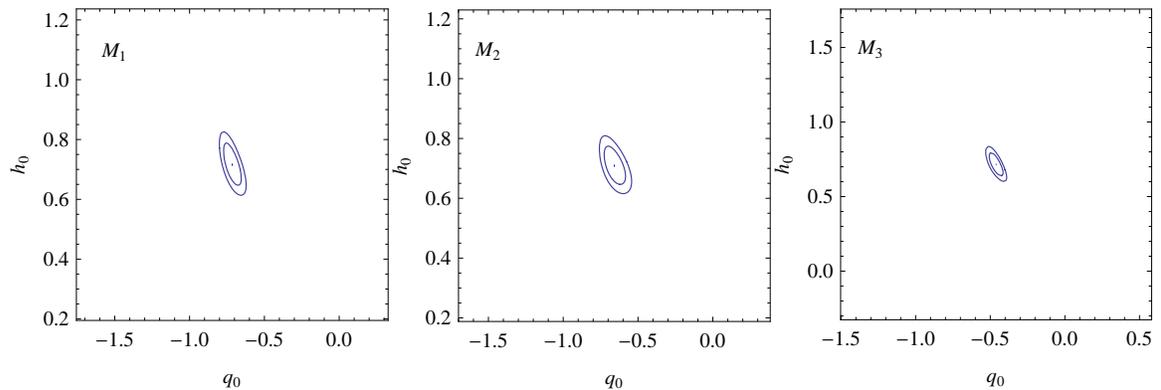}
\caption{The contour plots of $q_0-h_0$ with $1\sigma$ and $2\sigma$
regions. The left, center and right panels correspond to $M_1$
($q(z)=q_0+q_1z/(1+z)$), $M_2$ ($j=j_0$) and $M_3$(expansion of
luminosity distance) respectively, where the dots denote the best
fit values of the parameters.}\label{fig:con_q_h}
\end{figure}

The $1\sigma$ and $2\sigma$ contour plots in $q_0-j_0$ and $q_0-h_0$
planes are plotted in Fig. \ref{fig:con_q_j} and Fig.
\ref{fig:con_q_h}. In $q_0-j_0$ contour plots \ref{fig:con_q_j}, the
$\Lambda$CDM model result is included as a short line segment
denoting $1\sigma$ interval where the dot denotes the best fit value. It can be seen from Fig.
\ref{fig:con_q_j}, the best fit $\Lambda$CDM model is out of the
range of $1\sigma$ region in $M_1$ and $2\sigma$ region in $M_3$,
but it is in the range of $1\sigma$ region in $M_2$. It means that
the best fit $\Lambda$CDM model is compatible at $1\sigma$ and
$2\sigma$ confidence levels with $M_2$ and $M_1$ respectively. The
results are compatible with that of Ref. \cite{ref:kinematic2}
except the model of expansion of luminosity distance. The results
imply a universe with $j_0<-1$ from $M_1$ and $M_2$.

\section{Conclusions}

In this paper,  kinematic models are constrained by recent cosmic
observations which include SN Ia, BAO and observational Hubble data
(OHD). Here, we give three simple examples: Taylor expansion of
$q(a)$ at present $a_0=1$ ($M_1$), constant jerk parameter $j=j_0$
($M_2$) and expansion of luminosity distance ($M_3$). All of the
models give the same prediction of the evolution of the universe
which is undergoing accelerated expansion at current and experiences
a transition from decelerated expansion to accelerated expansion.
But the transition redshift depends on the concrete parameterized
forms of the models. The best fit values of the parameter of  $M_1$
and $M_2$ predict the transition redshift is about $z_t\sim 0.6$.
However, $M_3$ predict a large transition redshift $z_t\sim 1$. The
$\chi^2/dof$ imply the same goodness of the models. But, from
evolution curves of deceleration parameter as plotted in Fig.
\ref{fig:qz} and the knowledges of the history of the universe, one
found that $M_1$ and $M_2$ are better than $M_3$ for the badness of
behavior at high redshift of $M_3$, say $z>1$.

The $1\sigma$ and $2\sigma$ confidence contours in $q_0-j_0$ and
$q_0-h_0$ planes are plotted in Fig. \ref{fig:con_q_j} and Fig.
\ref{fig:con_q_h} where the best fit value of $\Lambda$CDM model is
denoted by a line segment, $1\sigma$ interval. From the Fig.
\ref{fig:con_q_j}, one can find that $M_1$ and $M_2$ are compatible
with $\Lambda$CDM models at $2\sigma$ and $1\sigma$ confidence
levels respectively. $M_3$ is not compatible with $\Lambda$CDM model
at $2\sigma$ confidence level. For its badness of the evolution
history of model $M_3$ at high redshift. One can not take it
seriously and treat it as a unreliable model. Then, one can conclude
that the cosmic data favors the model having the value $j_0<-1$ for
the large parts of the confidence regions is under the line
$j_0=-1$. It is compatible with the conclusion obtained in Ref.
\cite{ref:kinematic2} .

\acknowledgements{This work is supported by NSF (10703001), SRFDP
(20070141034) of P.R. China.}

\appendix

\section{Models and useful relations}\label{sec:append}

In this appendix, one can find the analytical expressions between
the Hubble parameter $H(z)$, deceleration parameter $q(z)$ and jerk
parameter $j(z)$.

$\bf M_1$
\begin{equation}
q(z)=q_0+q_1\frac{z}{1+z}
\end{equation}
\begin{equation}
H(z)=H_0\left(1+z\right)^{1+q_0+q_1}\exp\left(-\frac{q_1
z}{1+z}\right)
\end{equation}
\begin{equation}
j(z)=-q_0-\frac{q_1 z}{1+z}-(1+z) \left(-\frac{q_1
z}{(1+z)^2}+\frac{q_1}{1+z}\right)-2 \left(q_0+\frac{q_1
z}{1+z}\right)^{2}
\end{equation}

$\bf M_2$
\begin{equation}
 H(z)= H_0 [ c_1(1+z)^{\alpha_1} + c_2(1+z)^{\alpha_2} ]^{\frac{1}{2}}
\label{M3huble}
\end{equation}
\begin{equation}
q(z)= \frac{c_1(1+z)^{\alpha_1}(\frac{\alpha_1}2-1) +
c_2(1+z)^{\alpha_2}(\frac{\alpha_2}2-1)}{c_1(1+z)^{\alpha_1}+c_2(1+z)^{\alpha_2}}
\label{M3_q}
\end{equation}
\begin{equation}
 j(z)=j_0
\end{equation}
\begin{equation}
 z_t= \left[ - \frac{c_2}{c_1}\frac{\alpha_2-2}{\alpha_1-2}\right]^
 \frac{1}{\alpha_1-\alpha_2}-1
 \label{M3zt}
\end{equation}
where
\begin{equation}
 \alpha_{1,2}=\frac{3}{2}\pm\sqrt{\frac{9}{4}-2(1+j_0)}
 \label{M3alpha}
\end{equation}
\begin{equation}
c_1=\frac{2(1+q_0)-\alpha_2}{\alpha_1-\alpha_2} \quad {\rm and}\quad
c_2=1-c_1
\end{equation}
From (\ref{M3alpha}) we see that $j_0<\frac{1}8$.

{$\bf M_3$} -- defined by the expanded luminosity distance Eq.
(\ref{eq:dL_exp}), $d_L(z)=\frac{c}{H_0} \left( z+ Az^2+ Bz^3
\right)$, where $A=(1-q_0)/2$ and $B=-(1-q_0-3q_0^2-j_0)/6$.
\begin{equation}
  H(z)= H_0 \left[ \frac{(1 + z)^2}
    {1+2Az+(A+3B)z^2+2Bz^3} \right]
\end{equation}
\begin{equation}
  q(z)= \frac{1-2A-2(A+3B)z-(A+9B)z^2-2Bz^3}{1+2Az+(A+3B)z^2+2Bz^3}
  \label{M4_q}
\end{equation}
\begin{equation}
  j(z)= -\left[ q + 2q^2 + (1+z) q^{\prime} \right]
\end{equation}
\begin{equation}
  z_t: {\rm the \; real \; root \; of \;} \;
  1-2A-2(A+3B)z_t-(A+9B)z_t^2-2Bz_t^3
\end{equation}

{\bf $\bf \Lambda$CDM}, $\Omega_m+\Omega_\Lambda=1$
\begin{equation}
 H(z)= H_0 \left[ \Omega_m (1+z)^3 + (1-\Omega_m)\right]^{\frac{1}{2}}
\end{equation}
\begin{equation}
 q(z)= \left[(1+z)^3-2(1/\Omega_m-1)\right]
 / \left[2(1+z)^3+2(1/\Omega_m-1)\right]
\label{lcdm_q}
\end{equation}
\begin{equation}
 j(z)= -1
\end{equation}
\begin{equation}
 z_t= [2(\Omega_m^{-1}-1)]^{\frac{1}3}-1
\end{equation}
Note that the expressions for $\Lambda$CDM can be easily obtained
from (\ref{M3huble}-\ref{M3zt}), putting $j_0=-1$ into
(\ref{M3alpha}).

\end{document}